# Why Can the Image of the City be Formed?

Bin Jiang

Department of Technology and Built Environment, Division of Geomatics
University of Gävle, SE-801 76 Gävle, Sweden
Email: bin.jiang@hig.se

The aim of this short paper is to summarize my recent preprint (arXiv:1209.1112), in which I proposed a novel and probably controversial view about cognitive mapping; that is, the image of the city out of the underlying scaling of city artifacts or locations. The scaling refers to a recurring structure of far more small things than large ones. Five decades of research on the image of the city (Lynch 1960) or cognitive mapping in general has focused almost exclusively on internal representation about how the city is represented internally in human minds, leaving the external representation – or the city itself – out of the scope of scientific investigation on cognitive mapping. I claimed in the preprint that it is primarily the city itself – or, more precisely, its intrinsic scaling property – that makes the city imageable or legible. Put another way, if the city were devoid of the scaling property, it would be very hard to form the image of the city internally. In this paper, I attempt to further clarify the central argument, and identify some possible areas of misunderstanding for readers.

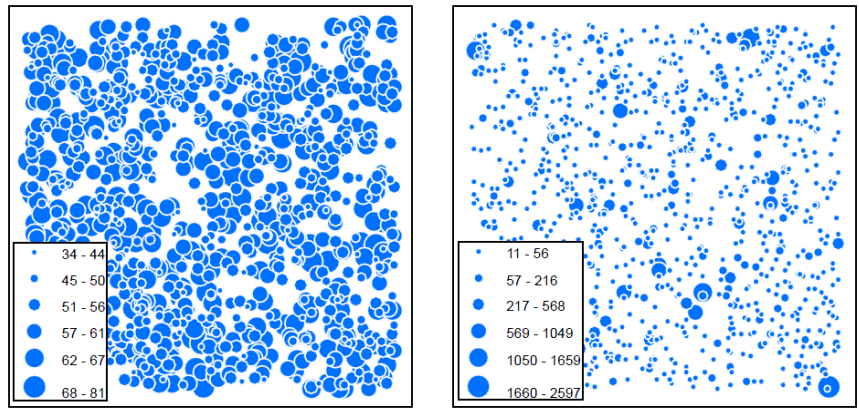

Figure 1: (Color online) Two point patterns generated from (a) the normal distribution dataset and (b) the heavy-tailed distribution dataset

There are two perspectives on cognitive mapping research: one is the human perspective that focuses on how human beings perceive a city and form the image of the city internally; the other is the city perspective as to what traits a city has to help form the image of the city. However, the literature has focused almost exclusively on the first perspective with probably two exceptions (Lynch 1960, and Haken and Portugali 2003). Lynch (1960) formulated the theory on the image of the city, and coined two terms to characterize cities and city artifacts: imageability and legibility. He further extracted five geometric or visual elements: paths, edges, districts, nodes, and landmarks with high imageability or legibility. Haken and Portugali (2003) developed an information view, which argued that it is information, in particular semantic information that cities or city artifacts have, that makes the cities or city artifacts imageable or distinguishable.

My proposal is in line with the two above-mentioned views. I argue that it is the scaling of city artifacts or locations that makes the city imageable, or the city artifacts or locations distinguishable from the vast majority of other redundant city artifacts. I further argue that scaling is the first and foremost effect, without which the image of the city can hardly be formed. Let me highlight my central argument by asking the following question: given the two patterns shown in Figure 1 (c.f. Page 1 or Figure 8 of the preprint), which pattern looks more like a city or a system of cities?

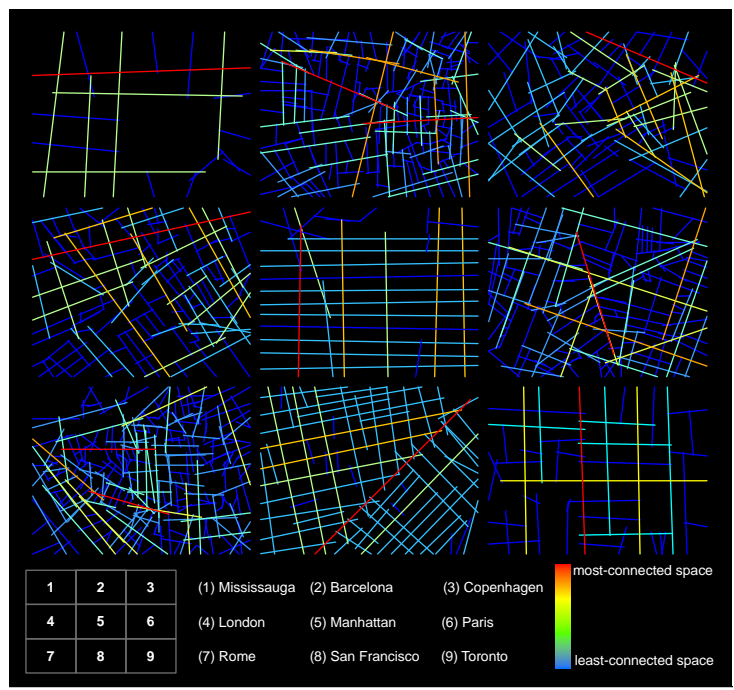

Figure 2: (Color online) Topology of typical street patterns

The reader may respond by saying that neither pattern looks like a city, since a city is full of streets and buildings, neither of which are shown in the patterns. Streets and buildings are what the city looks like, not its essence or its order under the surface. The essence of the city, or the order under the surface, is the scaling property: far more small things than large ones. A reader who captures this essence of the city is more likely to say that the right-hand pattern looks more like a city. The key point of my argument is that the reason why the image of the city can be generated in human minds is to do with mental processing, but human internal representation is not the primary driver; the primary driver is the scaling property of the city artifacts or locations. If the city were like the left-hand pattern, the image of the city would hardly be generated in human minds. To the best of my knowledge, none of the early literature or the spatial cognition literature has touched upon this fundamental issue.

To compute the image of the city, we simply rank all city artifacts according to their geometric, topological and/or semantic properties from the largest to the smallest (Jiang 2012b), and conduct the head/tail breaks classification (Jiang 2012a, Jiang 2012c). Eventually, the largest artifacts constitute the image of the city. In Figure 2 (c.f. Page 2 or Figure 6 in the preprint), the red lines (or the most connected lines) constitute the image of the cities. With Figure 2, note that the axial lines are treated as all city artifacts of the individual cities, and the coloring is based on the topological information, the degree of street connectivity, so to speak. It is important to remember that this study is *not* about axial line analysis or space syntax in general (Hillier and Hanson 1984). The axial lines are used as a proxy for city artifacts that demonstrate the scaling property; that is, far more less-connected lines than well-connected ones.

One might question how, without examining the internal representation or conducting human subject tests, it is possible to be sure that the red lines constitute the image of the city. It appears common sense that the largest are kept in our memory (Omer and Jiang 2010). The largest are in terms of geometry (e.g., size, shape, and/or orientation), topology (connectivity), semantics (meaning), and/or combinations of these. For example, the longest streets are surely in the image of the city, the most connected streets are surely in our memory, and one short street in which a country's queen was assassinated is surely kept in the memory of its residents. Note that the largest are in the context of scaling, rather than for things that are normally distributed.

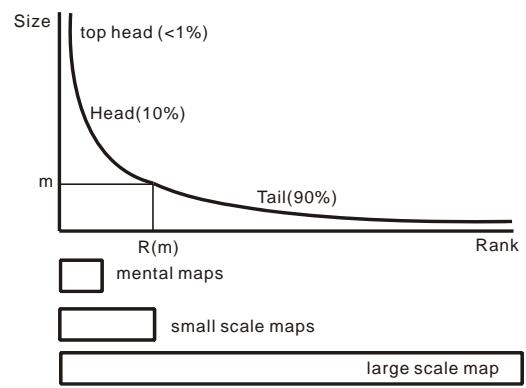

Figure 3: Unification of cognitive and cartographic mapping under the same framework of scaling

The above process of computing the image of the city is much like that of map generalization (Jiang et al. 2012). This is illustrated in Figure 3 (or Figure 9 in the preprint), in which large-scale and small-scale maps, as well as mental maps, correspond to different parts of the scaling curve. Note that the scaling curve should be understood in a recursive manner; that is, the head part can be re-stretched to form another scaling curve for multiple times. This is the true sense of the recurring structure of far more small things than large ones.

In summary, this paper proposes a novel and probably controversial view about cognitive mapping, stressing that the external scaling property is the primary driving force for forming the image of the city. Under the same scaling law, cognitive and cartographic mapping can be unified as a ranking and selection process from massive amount of city artifacts or locations by considering geometric, topological and/or semantic attributes. This is particularly true under the circumstance where the increasing amounts of social media data (e.g., Flickr and Twitter) can be used for cartographic and cognitive mapping. Scaling might be formulated as a law of geography.

## References

(Note: The references are apparently biased, and one should instead refer to Jiang (2012c) and the references therein)

Haken H. and Portugali J. (2003), The face of the city is its information, *Journal of Environmental Psychology*, 23, 385 – 408.

Hillier B. and Hanson J. (1984), *The Social Logic of Space*, Cambridge University Press: Cambridge.

Jiang B. (2012a), Head/tail breaks: A new classification scheme for data with a heavy-tailed distribution, *The Professional Geographer*, xx(x), xx – xx.

Jiang B. (2012b), Computing the image of the city, In: Campagna M., De Montis A., Isola F., Lai S., Pira C. and Zoppi C. (editors, 2012), *Planning Support Tools: Policy analysis, implementation and evaluation, Proceedings of the 7th Int. conf. on Informatics and Urban and Regional Planning INPUT 2012*, 111-121.

Jiang B. (2012c), The image of the city out of the underlying scaling of city artifacts or locations, Preprint: http://arxiv.org/abs/1209.1112.

Jiang B., Liu X. and Jia T. (2012), Scaling of geographic space as a universal rule for map generalization, Preprint: http://arxiv.org/abs/1102.1561

Lynch K. (1960), *The Image of the City*, The MIT Press: Cambridge, Massachusetts.

Omer I. and Jiang B. (2010), Imageability and topological eccentricity of urban streets, in: Jiang B. and Yao X. (editors), *Geospatial Analysis and Modeling of Urban Structure and Dynamics*, Springer: Berlin, 163-176.